# Ferromagnetic Phase Transition of DPPH Induced by a Helical Magnetic Field

**Emmanouil Markoulakis [1], John Chatzakis [1], Antonios Konstantaras [1], Iraklis Rigakis [1] and Emmanuel Antonidakis [1]**

[1] School of Engineering, Department of Electronic Engineering, Computer Technology Informatics & Electronic Devices Laboratory, Hellenic Mediterranean University, Romanou 3, Chania, 73133, Greece; markoul@hmu.gr

**Abstract**

We report the results and unique instrument configuration of a novel experiment in which we successfully phase transitioned a DPPH sample from its natural paramagnetic state, essentially a non-magnetic material, to a ferromagnetic state at room temperature using a specifically applied helical flux magnetic field. The DPPH sample (2,2-diphenyl-1-picrylhydrazyl) remained ferromagnetic at least one hour after the experiment, suggesting that a transformation in the material was induced by the external field rather than being merely a magnetic phase transition observed only during the experiment. The external magnetic field used had a helical pitch angle of approximately 54.7°, known mathematically as the Magic Angle, to the +z-axis, thus to the normal S to N external field's magnetic moment vector. Based on the phenomenology of the experiment and results, we suggest that this specific magic angle, which corresponds also to the known quantization precession spin angle of free electrons under a homogeneous straight flux magnetic field, potentially enhances the percentage of unpaired valence electrons within the DPPH material that align in parallel with the external applied field. Typically, in paramagnetic materials, the distribution of unpaired electrons' quantum spins relative to an external field is nearly random, exhibiting roughly a 50% chance of either parallel or antiparallel alignment. Only a slight majority preference exists in one alignment direction due to the Boltzmann thermal distribution, which contributes to the paramagnetic nature of these materials. We measured the induced ferromagnetism of our DPPH sample and found its relative magnetic permeability to be an abnormally thousand-fold increased in decimal value of $\mu \approx 1.4$, compared to its normal typical paramagnetic value of 1.0001 for this material.

---

## 1. Introduction

It is well known in theory but also experiments like the classical Stern-Gerlach [1][2] experiment that of the binary quantization of the intrinsic magnetic spin of the electrons in space either spin up or spin down. Also, an equally important observation from quantum mechanics is that we cannot control this quantum spin of free electrons [3][4] which have equal statistical random 50% probability to end up either aligned parallel to an external homogeneous B field's magnetic moment direction or antiparallel at an opposite 180° angle and that the intrinsic magnetic moment vectors of these free electrons or else called quantum spin of the electrons are precessing around an external magnetic B field's vectors as shown in this animation https://tinyurl.com/4uw27ax5 as well as in **Fig.1**. This spin precession frequency is called Larmor frequency and varies with the strength of the applied external magnetic B field with the relation $\omega_s=\gamma B$ where $\gamma$ is the gyromagnetic ratio intrinsic constant of the electron and that for one Tesla external field this corresponds for the free electron to about 28 GHz. However that what is not very widely known is that the precession spin angle is always fixed *see* **Fig.1,** has always a fixed angle of approximately |54.7°| to the +z–axis or its complementary of |125.3°| to the -z–axis depending the random parallel or antiparallel alignment of the electron to the external magnetic B field, spin up or spin down. This specific quantization intrinsic to the free electron, spin angle [5][6] remains fixed independent to the strength B of the external magnetic B field as shown in the below quantum mechanical analysis:

From atomic structure we know an electron can only have an intrinsic spin vector of $s=½$. From quantum mechanics the magnitude of the spin vector $|s|$ is,

$$| s |= \sqrt{s(s+1)}\hbar = \sqrt{3/4}\hbar \, . \qquad (1)$$

Where $\hbar$ is the reduced Planck constant. The radius angle of the cones of uncertainty, **Fig. 1**, calculates therefore as shown in equation (2),

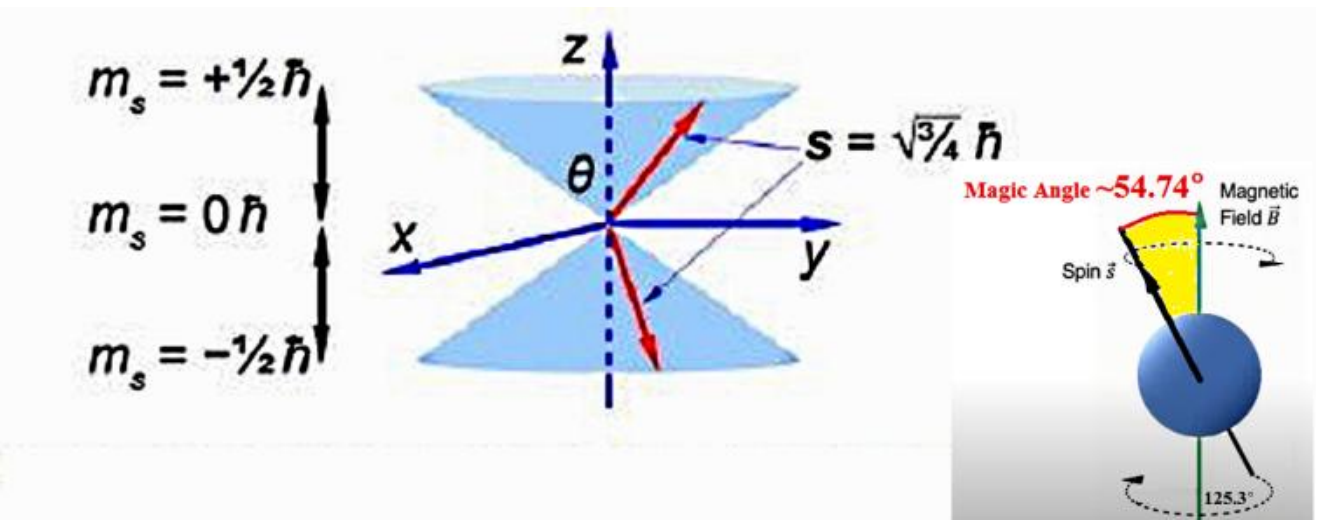

**Figure 1.** The two allowed orientations of the *s*=½ spin angular momentum vector.

$$\begin{pmatrix} m_s = +\frac{1}{2} \theta = \cos^{-1}\left( \frac{1/2\hbar}{\sqrt{3/4}\hbar} \right) \approx 54.7^{\circ} \\ \\ m_s = -\frac{1}{2} \theta = \cos^{-1}\left( \frac{-1/2\hbar}{\sqrt{3/4}\hbar} \right) \approx 125.3^{\circ} \end{pmatrix} . \qquad (2)$$

This absolute quantization angle of ∣54.7°∣ to both the ±z–axis (complementary is at 125.3°) is therefore fundamental and intrinsic to the free electron and coincides with the so called $\theta_m \approx 54.7356°$ Magic Angle [7][8] found in mathematics, defined in equation (3) as:

$$\theta_{\mathrm{m}} = \arctan\sqrt{2} \approx 0.95532 rad \approx 54.74^{\circ} , \qquad (3)$$

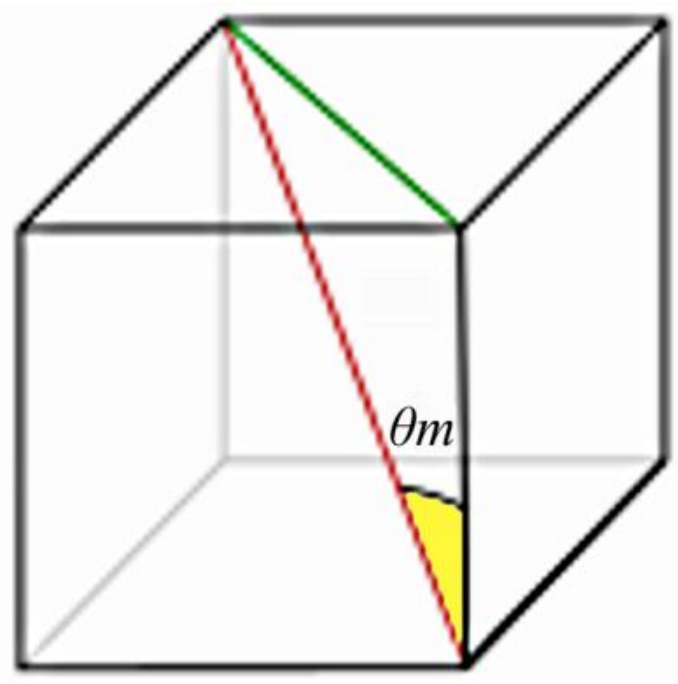

**Figure 2.** Magic Angle inside the cube.

and embedded as a fixed angle in the geometry of the cube, **Fig.2**.

However, besides the quantum mechanical description of the magic angle intrinsic characteristic of the electron, as far as we know there is no any physical interpretation existing for this angle. The authors believe that the magic angle is an intrinsic property of the charge of the electron as also theorized in our previous research [9].

We theorize herein [9], indirectly proving our assumption, that an externally applied helical magnetic field with this same exact angle could potentially interfere with the charge of the electron and its magnetic dipole moment and in some extend artificially control its quantum spin direction to align parallel to the external magnetic B field vector. The concept is similar with turning in or out a mechanical screw to the matching threads in the hole.

To prove this, a novel experiment was devised alternative to a modified Stern-Gerlach experiment option without the need of a high-vacuum environment for the free electrons. We will describe next this experiment setup and results obtained and discuss the implications and practical applications of our findings but also some fundamental and possible new physics from our research.

## 2. Materials and Methods

### *2.1. DPPH Sample*

In order to emulate free electrons in our experiment without the need of a high-vacuum environment as in the case of a modified SG-experiment, we have chosen to use DPPH [10][11][12][13], 2,2-diphenyl-1-picrylhydrazyl, see **Fig.3(c)**, an organic chemical compound of dark colored crystalline powder. It can be characterized as a macroscopic quantum spin emulator material since its free radical molecules have a single loosely bound unpaired electron per molecule (every 41 atoms) therefore emulating close enough free electrons in a normal experiment environment on air without the need to use a high-vacuum Its unique properties and g-factor of ≈2.0036 which is very close to that of a free electron is the reason why it is used in Electron Paramagnetic Resonance (EPR) spectroscopy as a reference calibration material. Our lab purchased 1g of DPPH powder from which a 36mg sample was used in the experiment.The DPPH powder material is normally paramagnetic, having a typical relative magnetic permeability of $\mu_r \approx 1.0001$.

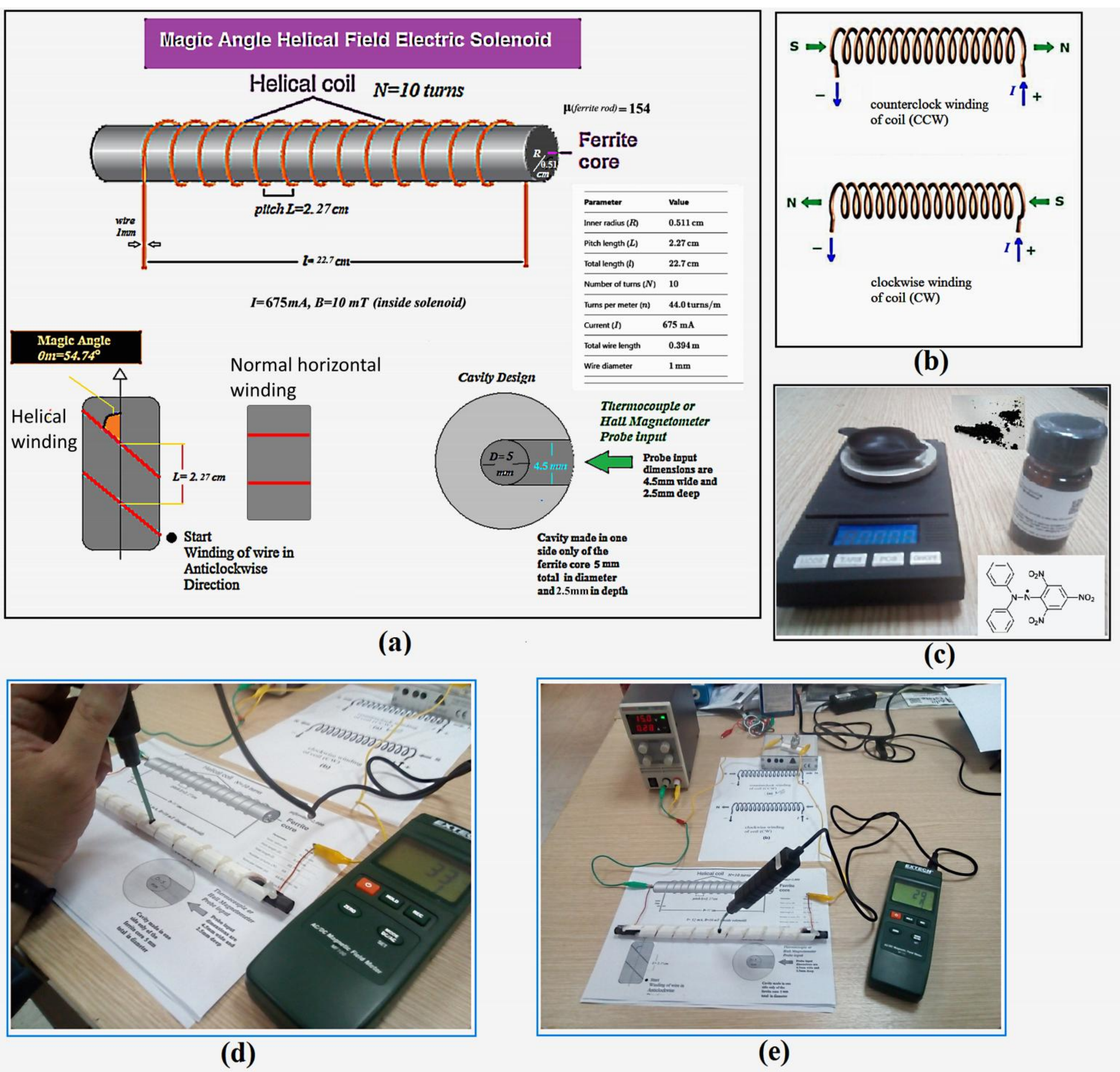

**Figure 3.** **(a)** Blueprint design of the special constructed prototype Magic Angle Helical Field electrical solenoid with a MnZn ferrite rod core. A standalone larger in size version of the blueprint can be found here, https://tinyurl.com/bdh89cze. **(b)** Sense or else polarity of the magnetic field N-S created in the solenoid by a d.c. electric current $I$ passing through the coil for the case of a counterclockwise (CCW) winding of the coil or clockwise (CW) winding as shown accordingly. In the experiment the first case shown (top illustration) was used that of a CCW winding of the coil since this sense matches the intrinsic chirality of a free left-handed electron. **(c)** The purchased 1g of DPPH. We used only a sample of 36mg in the experiment we have weighted with a 1mg resolution digital scale. **(d)&(e)** Actual photographs of the experiment setup and measurements with the Hall sensor Magnetometer Extech MF-100 model.

### 2.2. Magic Angle Helical Field Solenoid

In **Fig.3(a)** we see the blueprint of the design of the prototype solenoid used in the experiment. A standalone larger in size version of the blueprint can be found here, https://tinyurl.com/bdh89cze. A calculated pitch step length (see Appendix I in supplementary material) of the winding turns of the coil at $L$=2.27 cm for a total length of the solenoid at *22.7* cm and *10* turns shown in the bottom left corner of the blueprint, results to the magic angle (i.e. ≈54.74°) helical winding of the solenoid along its length axis that generates inside the electric solenoid when a d.c. current $I$ passes through, a magnetic field with a helical flux of the same pitch magic angle.

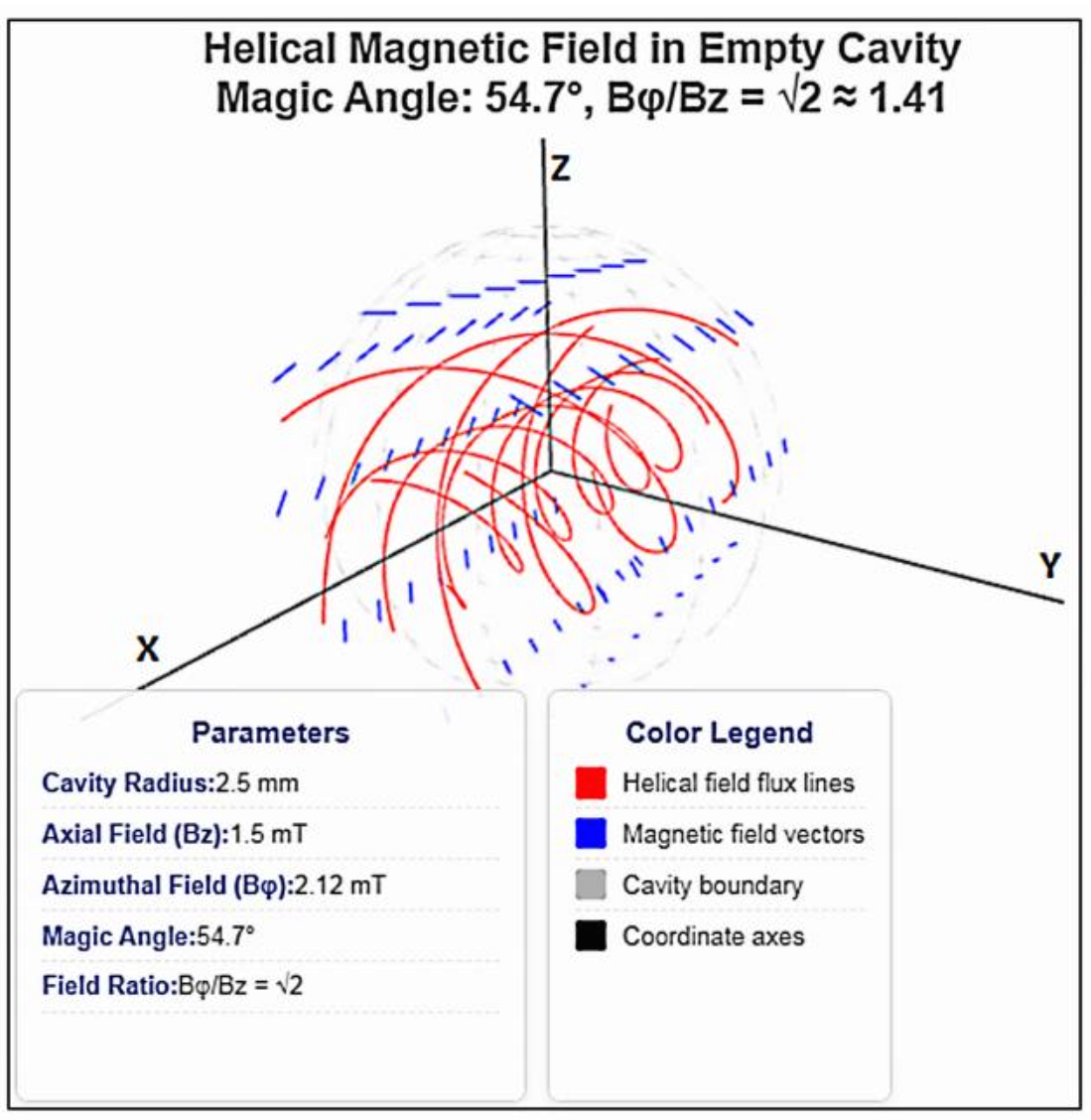

**Figure 4.** Cartesian 3D plot of the helical field inside the cavity.

In **Fig.4** we see a simulated three-dimensional plot of the helical magic angle magnetic field inside the empty cavity.

Notice here a very important detail and key characteristic of the design of the prototype solenoid that differentiates it from a normal wound solenoid. For a standard, "ideal" solenoid, the coil (wire) turns are considered perpendicular to the central axis of the solenoid along its length (i.e. normal "horizontal winding"). In this case, each turn is essentially a circular loop, and the superposition of the magnetic fields from all these loops results in a uniform, axial straight flux magnetic field close to the center inside the solenoid. However, when the coil is wound at an extreme, as in our case, skewed angle (not perpendicular to the axis along the length of the solenoid), see **Fig.3(a)** bottom left corner, the current path has two components: A component that circles the axis, contributing to the axial magnetic field. Secondly, a component that runs parallel to the axis, which creates an azimuthal (circular) magnetic field around the central line.

This combination of an axial field $B_z$ and an azimuthal field $B_\Phi$ gives the overall magnetic flux a corkscrew or helical shape inside the solenoid. The specific skew magic angle of approximately $\theta_m \approx 54.74°$ the coil is wound, is a critical value for creating a helical field inside our prototype solenoid, and does influence the ratio of the azimuthal $B_\Phi$ to axial $B_z$ field components as shown in equation (4):

$$\tan(\theta_m) = \frac{B_\phi}{B_z} = \sqrt{2} \approx 1.41 \Rightarrow \theta_m = \arctan(\sqrt{2}) \approx 54.74^{\circ}. \quad (4)$$

A more analytic derivation of the field components can be found in Appendix II in supplementary material.

The key factor is simply that the winding is not perpendicular to the length axis. The flux lines will twist as they travel down the length inside the solenoid, matching the path of the current running inside the helical coil. This principle method of creating helical field is used in specific applications, such as in nuclear fusion stellarators. When introducing a ferrite rod core inside the solenoid the magnetic flux inside the solenoid and ferrite core will still be helical, preserving the original flux geometry of the air helical wound solenoid but the presence of the ferrite core will significantly enhance and concentrate this helical flux.

A CNC drilled into the MnZn ferrite rod core at the center, semi-spherical cavity and a channel to insert the magnetometer probe sensor was made as shown at the bottom-right corner of the blueprint, see **Fig.3(a)** and at the supplemetary material demonstration example. This cavity has 5mm cross-section and is *2.5* mm deep. The volume of the cavity was calculated that would be filled up to the rim when only a 36mg sample of the DPPH powder is used and inserted inside the cavity. The MnZn ferrite rod core used is *28* cm long and *10.2* mm in diameter.

Besides, the physical dimensions and operational electric and magnetic characteristics of the solenoid shown in detail in **Fig.3(a)** but also **Fig.3(b)** showing the resulting magnetic polarity N-S of the solenoid (we used the counterclockwise CCW sense winding of the solenoid, see top illustration in Fig.3b), it is very important to calculate the effective relative magnetic permeability $\mu_r$*(ferrite-rod)* of the MnZn ferrite rod core of the solenoid. For that, we *measured* with an RLC Bridge very accurately at 1KHz a.c. current the inductance of the prototype solenoid with the ferrite core inserted, being $L_{ind}=7\mu H$ and then solving equation (5) for $\mu_r$ we calculated the relative magnetic permeability:

$$L_{ind} = \frac{\mu_r \mu_0 N^2 A}{\ell} \quad \begin{aligned} \ell &= \text{length of solenoid} \\ A &= \text{cross-sectional area} \end{aligned} . \tag{5}$$

Where N=10 is the number of turns of the prototype solenoid and $\mu_0$ the permeability of free space giving the effective relative magnetic permeability of the MnZn ferrite rod core at $\mu_r$(ferrite-rod)≈154 for room ambient temperature (i.e. 300K). This last result is important in order to proceed with our experiment. In **Fig.3(d)&(e)** we see the actual experiment setup and measurements with a Hall sensor magnetometer. The thin green PCB board probe of the magnetometer is shown to be inserted inside the CNC drilled channel into the ferrite rod and cavity. A normal d.c. power supply 30V/5A was utilized in series with a current limiter ohmic resistance of R=6Ω. A white 3D printed winding guide shown was created and used to ensure the precise at the magic angle 54.74° uniform helical winding of the prototype electric solenoid. The 3D printing files are available in the supplementary material here https://tinyurl.com/34xdt5wr.

*2.3. Experiment Procedure*

In total over 200 field measurements were taken over various values of input d.c. current *I* to the in-series connected prototype solenoid with an *R=6Ω* current limiting ohmic resistance. Each measurement was repeated five times *(i.e. five runs of the experiment)* and the obtained measured values were averaged. These measured values can be categorized into three sets of measurements taken, each set corresponding to a different discrete experiment setup as illustrated in **Fig.5**. As we can see the Hall magnetometer probe consists from a thin plastic PCB strip on which at its end tip on one of its two sides, the tiny ≈1mm$^2$ Hall sensor chip is located as shown in **Fig.5** with orange (or light color rectangle for B&W). The probe can be turned either side to the left or to the right, by hand so that the orange colored sensor chip faces either the empty air cavity see **Fig.5(b)** or is in physical contact with the MnZn ferrite core bulk. see **Fig.5(a)**. In the first case (a), a set of measurements is obtained where we measure the strength of the magnetic field B in contact with and inside the MnZn ferrite rod core of our prototype magic angle solenoid, previously measured having an effective relative magnetic permeability of $\mu_r$*(ferrite-rod)≈154.* In case (b) a different discrete set of measurements are taken of the magnetic field strength this time inside the empty air cavity. In the last case illustrated in **Fig.5(c)** at the bottom, a third set of measurements is obtained in which we measure the magnetic field B strength when the previously empty cavity is completely filled with our 36mg DPPH crystalline powder sample. Therefore, measuring the magnetization and magnetic field contribution of the DPPH sample under the applied magic angle helical magnetic field inside our prototype solenoid.

Next we will present in summary and discuss these results.

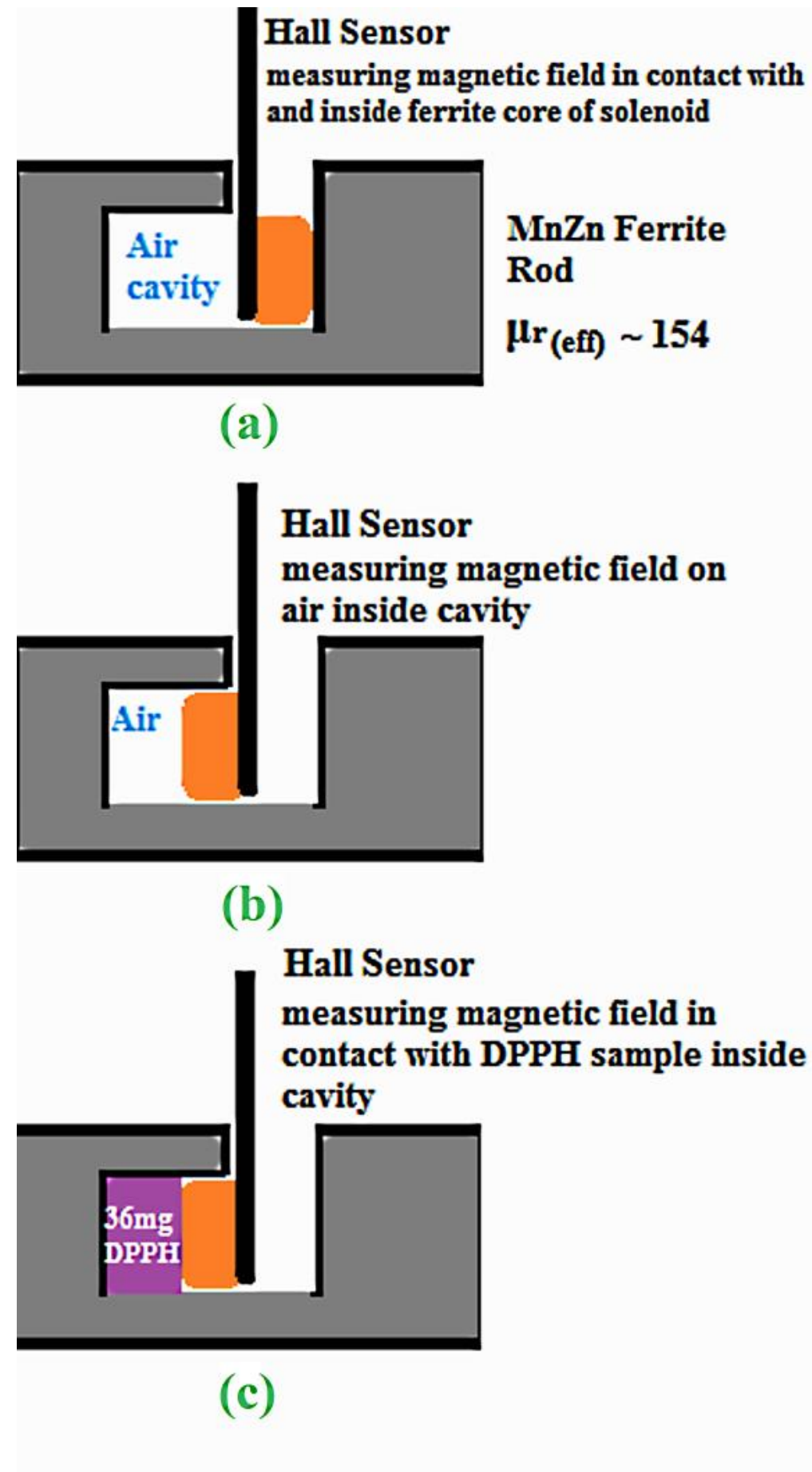

Figure 5. Three sets of measurements corresponding to (a), (b) and (c).

## 3. Results and Discussion

### *3.1. Measurement Data Summary*

The characteristic plot diagrams shown if **Fig.6** nicely summarizes all the important data from our measurements obtained during the five experiment runs we have done and by averaging all the measured values from different runs of the experiment. A statistical error bar of ±50μT is indicated for all the field measurements:

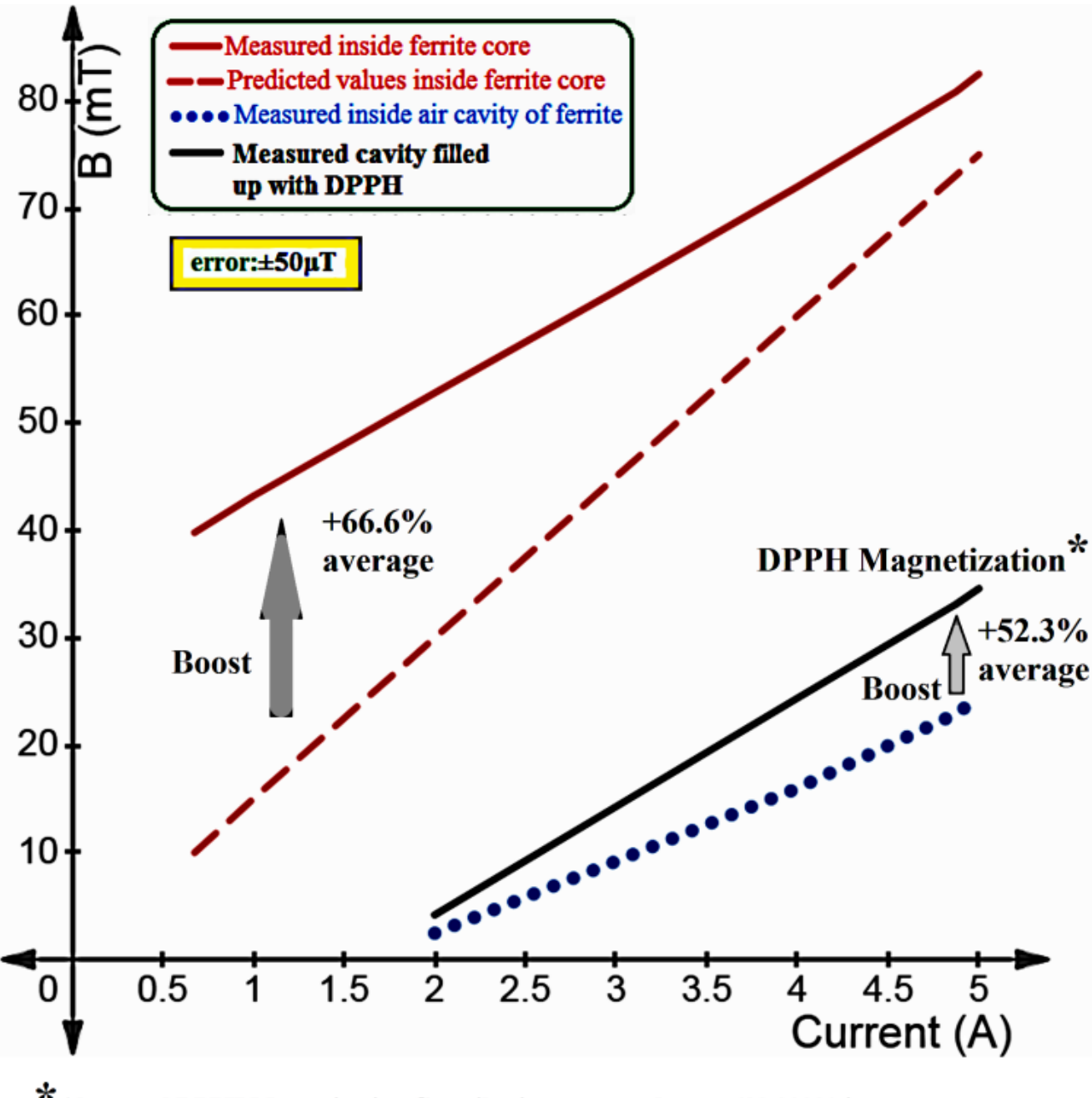

*Abnormal DPPH Magnetization Contribution measured at ≈ +632,000% in average.

**Figure 6.** Measured magnetization plots and magnetic field strength B vs. input d.c. current *I* inside the prototype helical solenoid for various experiment setups, i.e. inside ferrite core bulk, inside empty air cavity, and inside cavity filled with a 36mg DPPH sample.

The red dashed plot line (i.e. second line from top, shown in Fig.6) represents the theoretical predicted applied field strength values inside the MnZn ferrite core bulk material, without including any residual magnetization of the core, for a normal "horizontal windings" solenoid, see **Fig.3(a)**, with coil turns with the same pitch spacing and which has however the same inductance $L_{ind}=7\mu H$, number of turns and ferrite core $\mu_r\approx154$ with our magic pitch angle helical wound prototype solenoid and therefore both solenoids should effectively be near to equivalent, whereas the red solid line (i.e. first line on top) represents the actual measured magnetic field strength values inside the MnZn ferrite core bulk of our prototype. We observe a +66.6% in average boost and discrepancy of the actual measured values compared to the theoretical predicted. Nevertheless, we deem this observed discrepancy as not sufficient to draw any conclusions since this observed boost could be partially contributed to a residual magnetization of the core and the ferrite rod material is already ferrimagnetic and not paramagnetic and besides lacks the ability to emulate the quantum spin of free electrons in contrast to the DPPH which is an electrons' quantum spin macroscopic emulator material as discussed previously.

This is best demonstrated in the next pair of plot lines located at the bottom of graph in **Fig.6** comparing the actual measured values inside the drilled empty air cavity in the ferrite core, see blue-dotted line at the bottom of the graph, with the actual measured values when the cavity is filled completely with our 36mg DPPH powder sample, see black solid line (i.e. third line from top shown in Fig.6).

A surprisingly abnormal magnetization boost of +52.3% in average, was measured which should not be at all the case since DPPH is a paramagnetic material $\mu_r\approx1.0001$ and both bottom lines shown in Fig.6, in air the blue-dotted line and in DPPH the black solid line at the bottom, should be in any case practically tangent to each other.

In other words, the magnetization contribution of the DPPH material should normally be negligible and make no difference in the measurements at all compared to the empty air cavity.

The magnitude of the observed discrepancy is more emphasized and made even clearer in **Table.1** presenting a characteristic sample of our overall measurement data:

Table 1. Abnormal measured DPPH Magnetization using a Magic Angle Helical External Magnetic Field.

| **External Magnetic Field (mT)** | | **Predicted Normal DPPH Magnetization *$\mu_r \approx 1.0001$* Contribution (μT)** | **Measured Abnormal DPPH Magnetization Contribution (μT)** | **Discrepancy Measured Value From Theoretical Predicted** - how many times larger (approx.) | **New Measured Abnormal Value for Relative Magnetic Permiability of *DPPH $\mu_r=l+x$*** (typical value is *$\mu_r \approx I.0001$*) |
|---|---|---|---|---|---|
| **DPPH** | **Air** | | | | |
| **4.2** | 2.5 | 0.227 | 1700 | x7,488 | ***≈1.405*** |
| **14.3** | 9.2 | 0.772 | 5100 | x6,605 | ***≈1.357*** |
| **24.4** | 16 | 1.32 | 8400 | x6,363 | ***≈1.344*** |
| **33.1** | 22.9 | 1.79 | 10200 | x5,697 | ***≈1.308*** |
| **34.6** | 24.4 | 1.87 | 10200 | x5,454 | ***≈1.295*** |

Each row in **Table.1** is the averaged values from five runs of the experiment.

We observe already a very large discrepancy from the measurement data per row line corresponding each line of data to the same each time amount of d.c. input current *I* set in the experiment by controlling the power supply in Ampere units value, flowing inside the prototype solenoid's coil. The first and second column on the left is indicating accordingly the field strength measured with the cavity filled with DPPH and as empty on air. These two columns should have normally had the same values; in contrast we observe an enormous magnetization contribution to the field by the DPPH. So, for example see at first data row line and third column, although the magnetization contribution to the external field of the 36mg DPPH sample was calculated to be a negligible amount of 0.227 μT thus a fraction of a μT and therefore a negligible tiny contribution of the normally paramagnetic DPPH, contrary we observe in **Table.1** an enormous 1700μT (i.e. 1.7mT) contribution increase thus a x7,488 thousand-fold larger measured discrepancy. This discrepancy is consistent also for the other values in the table and in average the discrepancy measured is in the order of x6,320 rounded value or else +632,000%.

In the last right-most column of **Table.1** we calculate the corresponding abnormal new relative magnetic permeability of the exposed DPPH sample during the experiment with a value around *$\mu_r$(Abnormal DPPH)* ≈1.4 which suggests a ferromagnetic material. Therefore, this can be regarded and classified based on our experiment data as a "magnetic phase transition" of the naturally paramagnetic DPPH $\mu_r \approx 1.0001$ typical value, we observed during the experiment, induced by this novel helical magic angle external magnetic field generated inside our prototype solenoid. Notice also, in this last column in its title description cell, the written equation inside $\mu_r=1+\chi$ where $\chi$ is the calculated magnetic susceptibility of the exposed DPPH sample and actually is the decimal point numerical component of the relative magnetic permeability $\mu_r$ value.

However, going one step further our findings suggest that the induced ferromagnetism in our DPPH sample by the external field was not just a magnetic phase transition observed during the experiment but actually has potentially transformed the DPPH turning it from being paramagnetic to being slightly ferromagnetic (i.e. notice here the value of 1.4 magnetic permeability is still a small ferromagnetic value compared to a typical value of 1000 for example for iron).

This is supported by our observation that the DPPH sample demonstrated ferromagnetic behavior even after the experiment has finished and we turned OFF the power, even one hour after as shown in this video here, https://tinyurl.com/3s4m2j5m were we have emptied the cavity from the DPPH material and as we can see, it is still attracted mainly the lighter flakes of the material, by an Neodymium N42 grade permanent cylinder magnet. This is further supported by the hysteresis loop we obtained of the sample after its exposure, shown here https://tinyurl.com/y4nanb2m, and also in **Fig.7**:

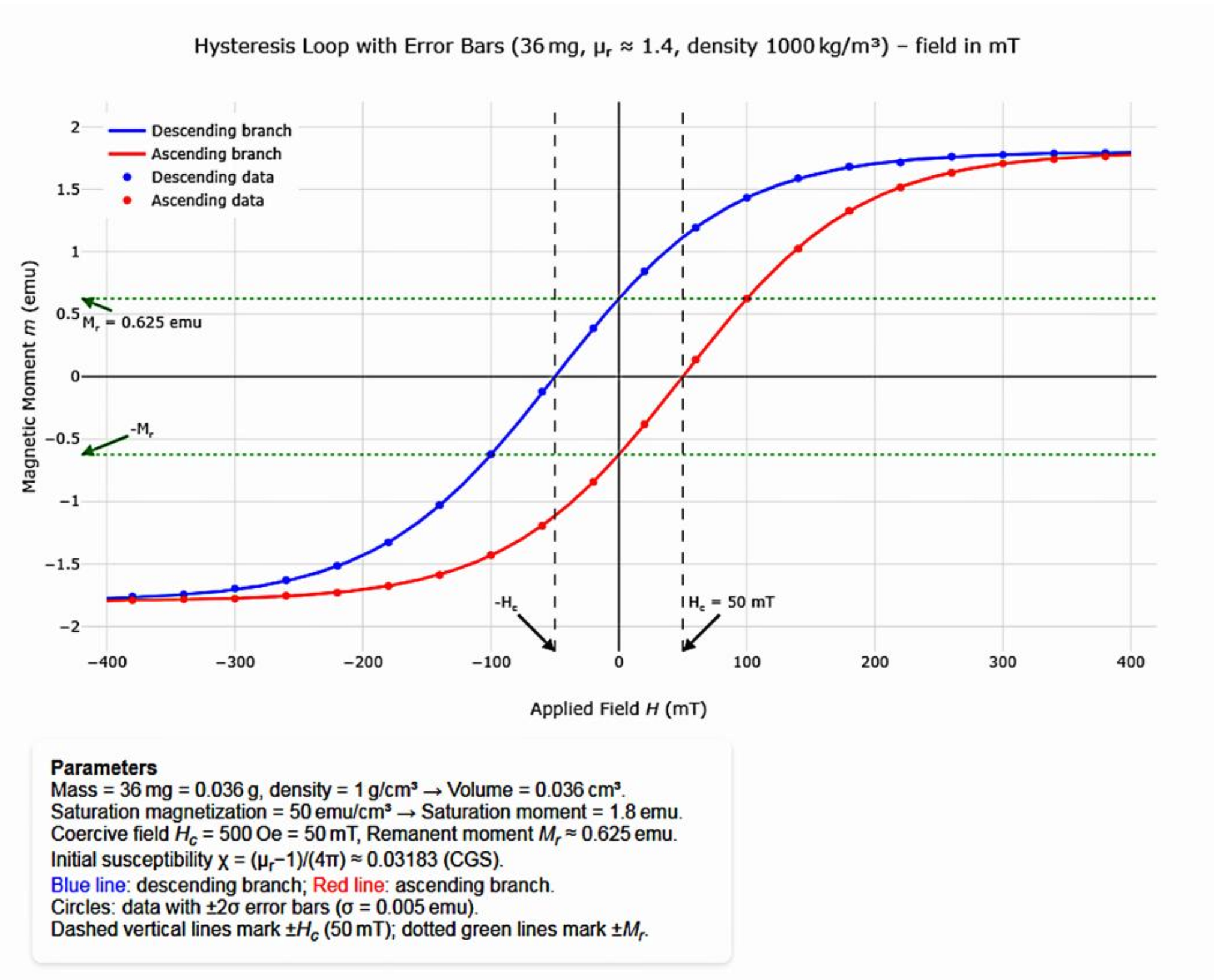

**Figure. 7** Hesteresis M-H loop of 36mg DPPH sample after its exposure.

Notice, the $M_r$ =0.625 emu (CGS units) value of Remanent (i.e. residual magnetization) magnetic moment $m$ shown in **Fig.7** corresponds to $0.625\times10^{-3}$A·m$^2$ in SI units for a direct comparison with the **Table.2** values shown in $\times10^{-8}$A·m$^2$ units thus a five orders of magnitude difference, and also that the saturation magnetic moment of our sample shown in **Fig.7** of 1.8 emu corresbonds to a magnetic flux density B of 62.83 mT. This weak ferromagnetism our sample demonstrated after its exposure to the experiment besides **Fig.7** data, is further supported by the statistical analysis of the results in Appendix III in the supplementary material and next **section 3.5** herein. Also, a magnetic-circuit analysis in Appendix IV.

### *3.2. Measurement Errors*

Apart of the accuracy statistical errors of the measuring apparatus, mainly the Hall sensor magnetometer given at ±50 μT error we deem as negligible and also because the macroscopic and simple nature of the experiment setup and procedure we therefore expect no meaningful persistent systemic errors in our experiment that could critically change the obtained results.

Additional measures were taken to avoid any systemic errors in the experiment preserving the experimental conditions all time like ambient temperature. Several runs of the experiment were performed at different days under the same always conditions. A stand was used to keep fixed the Hall sensor inserted inside the ferrite core without any manual intervention.

Maybe the most valid concern for a critical systemic error introduced was possible a contamination of our DPPH crystalline powder sample quantity of 36mg used, which was accurately measured with a 1mg accuracy digital scale and inserted inside the cavity. To avoid any contamination of the DPPH sample we have thoroughly cleansed the cavity with 99.9% pure alcohol and high pressurized air prior inserting our sample. After that the probe channel and cavity were always kept sealed with a tape wrapped around the ferrite core at the location of the cavity when in idle and no experiment was performed. Also, we never removed the Hall magnetometer probe which was fixed inside the cavity and hold in position by a stand, during all cavity measurements with the DPPH sample inserted, and a seal of tape was still applied around the aperture even with the inserted magnetometer probe to minimize exposure to the outside environment.

*3.3. Predicted normal magnetization of DPPH sample*

Because DPPH is naturally and normally a paramagnetic material with a typical relative magnetic permeability value of $\mu_r \approx 1.0001$ [14][15] it is predicted by theory but also empirically so far in the literature, to have a negligible magnetization and therefore also a tiny, negligible magnetic positive contribution to the applied helical magnetic field strength inside our prototype magic angle solenoid in the experiment. This tiny paramagnetic field strength contribution ΔB due to the magnetization $B_M$ of the 36mg DPPH sample was prior predicted to be a fraction of *one* $\mu T$ for external field strength $B_{ext}$ less than 10mT (i.e. 1mT=1000μT) and growing up to a maximum contribution of 10μT for an 100mT external field as shown in **Table.2**. Contrary, to the huge discrepancy found in our experiment results and previously described in **section 3.1.**. These nominal predicted values were calculated using the below described analysis.

The magnetization field contribution $\Delta B = B_M$ grows linearly with the applied external field $B_{ext}$, following equation (6) and are independent the mass and or volume of the DPPH sample:

$$\Delta B = B_M = \chi \cdot B_{\text{ext}} = \left(1.0 \times 10^{-4}\right) \cdot B_{\text{ext}} \qquad (6)$$

Where $\chi = 1 \times 10^{-4}$ is the typical magnetic susceptibility dimensionless value for DPPH. We see a summary of these theoretical predicted nominal results in **Table.2:**

**Table 2.** Predicted magnetic field normal contribution of DPPH sample per external applied field strength values.

| External Field (mT) | Magnetization M (A/m) | Total Magnetic Moment m ($\times 10^{-8}$ A·m$^2$) | Field Contribution ΔB (mT) |
|---|---|---|---|
| 1 | 0.080 | 0.29 | 0.0001 |
| 2 | 0.159 | 0.57 | 0.0002 |
| 5 | 0.398 | 1.43 | 0.0005 |
| 10 | 0.796 | 2.87 | 0.001 |
| 20 | 1.592 | 5.73 | 0.002 |
| 30 | 2.388 | 8.60 | 0.003 |
| 40 | 3.184 | 11.46 | 0.004 |
| 50 | 3.980 | 14.33 | 0.005 |
| 60 | 4.776 | 17.19 | 0.006 |
| 70 | 5.572 | 20.06 | 0.007 |
| 80 | 6.368 | 22.92 | 0.008 |
| 100 | 7.960 | 28.66 | 0.010 |

The magnetization *M* of the DPPH sample in the second column shown on **Table.2**, in A/m SI units can easily calculated using the equation (7),

$$M = \chi \cdot H = \chi \cdot \frac{B_{ext}}{\mu_0}, \qquad (7)$$

where $\mu_0$ is the magnetic permeability of free space constant, $\chi$ is the magnetic susceptibility of the DPPH material and *H* is the corresponding analogue field strength of the external applied field $B_{ext}$ inside the prototype solenoid's cavity when this field would have been applied instead in free space (i.e. vacuum). For linear materials like our paramagnetic DPPH or diamagnetic substances, where $\chi$ is small, this approximation of equation (7) is standard and widely applicable. Again, both field contribution $\Delta B$ and magnetization *M* of the DPPH sample do not depend on the mass or volume of the sample used in the experiment.

However, in the third column where we calculate the induced total magnetic dipole moment *m* in the 36mg DPPH sample, there, the magnetic moment is depended on the volume *V* of our sample used in the experiment (i.e. magnetic moment of DPPH sample scales with sample volume, mass-dependent). We remind here that by design, the DPPH sample volume should more or less coincide and be equal with the volume of the empty cavity inside the ferrite core assuming this cavity is completely filled up precisely by the 36mg DPPH powder sample.

Therefore using the equations (8) and (9),

$$m = M \cdot V \qquad (8)$$

$$V = \frac{\text{mass}}{\text{density}} = \frac{3.6 \times 10^{-5}}{1000} = 3.6 \times 10^{-8}\,\text{m}^3, \qquad (9)$$

where $m$ are the predicted total magnetic moment values of the DPPH 36mg sample calculated in the third column in **Table.2** in times $\times 10^{-8}$ A·m$^2$ SI units, and $M$ are the previously calculated in the second column magnetization values of the DPPH material in A/m SI units.

In equation (9) we used the conversion in Kg units, Mass=36mg=$3.6\times10^{-5}$kg and used a good density approximation for organic materials, thus, density of DPPH powder ≈ 1g/cm$^3$=1000kg/m$^3$.

*3.4. Control experiment with normal horizontal winding solenoid*

In order to farther test the validity of our results we repeated the experiment by constructing a normal "horizontal winding" solenoid with the same characteristics ferrite rod core. The second solenoid was designed to approximately produce the same field strength B values at the center inside the bulk of the core as the theoretical predicted values shown in **Fig.6** by the red dashed plot line (second line from top in Fig.6). This was achieved by using equation (10):

$$B = \frac{I \bullet L_{ind}}{NA}, \tag{10}$$

where $L_{ind}$ is the inductance of the solenoid, $I$ the electric current, $N$ the number of turns and $A$ is the cross-section area of the solenoid. This resulted to the design of the second "horizontal winding" solenoid with N=4 turns, 5cm in length, same cross-section 10.2 mm (diameter) as the helical prototype and having an inductance of $L_{ind}$≈5 μH. Again, the same cavity as in the helical prototype solenoid was drilled in the this second solenoid ferrite core ($\mu_r$≈154) at the center and channel for the Hall probe. The new measurements for the second solenoid and control experiment are summarized in **Fig.8**:

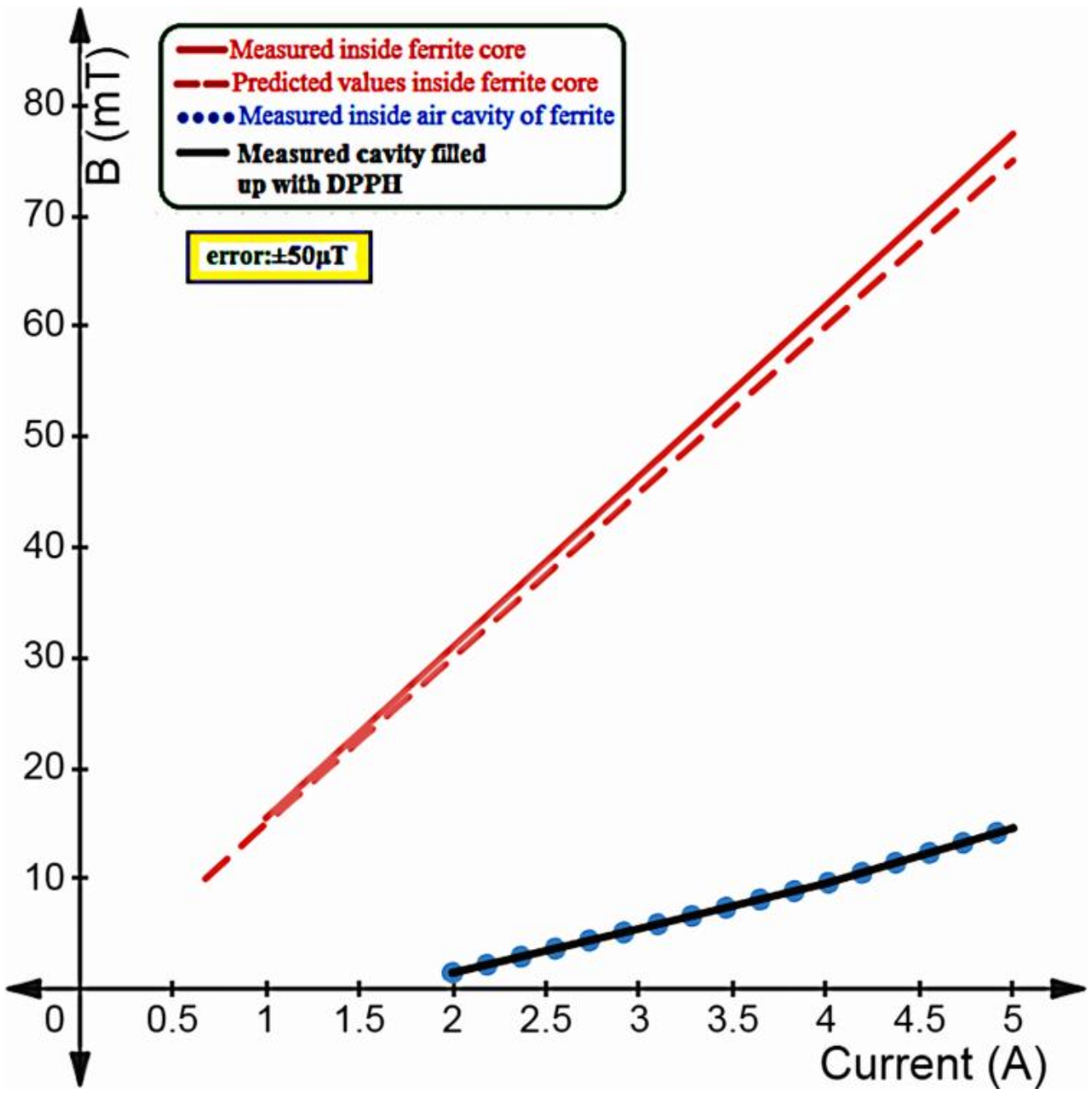

**Figure 8.** Measured magnetization plots and magnetic field strength B vs. input d.c. current $I$ inside the "horizontal winding" control experiment solenoid.

We observe from the results in **Fig.8** that the theoretical predicted values for inside the ferrite core (red dashed line, second from top) coincide nicely with the actual measured values (solid red line, first from top). Also, the actual measured plot lines for the empty air cavity and for the cavity filled with DPPH (blue dotted and solid black lines, at the bottom) are tangent to each other within the resolution capability of our measuring apparatus. Therefore, this validates the found discrepancy and our results previously presented for the prototype helical solenoid in section 3.1.

### 3.5. Example calculation of observed statistical new distribution of unbound electrons

*3.5.1 The Boltzmann Distribution in Normal Paramagnetic DPPH*

In a conventional paramagnetic material like DPPH, the alignment of unbound electron spins in an external magnetic field $B$ is governed by the Boltzmann distribution. The energy difference between a spin parallel (spin up, $m_s = +\frac{1}{2}$ ) and antiparallel (spin down, $m_s = -\frac{1}{2}$ ) is.

$$\Delta E = 2\mu_B B$$

where $\mu_B = 9.274 \times \frac{10^{-24}\,\mathrm{J}}{\mathrm{T}}$ is the Bohr magneton. The ratio of the number of spins parallel ( $N_+$) to antiparallel ($N_-$) is:

$$\frac{N_+}{N_-} = \exp\left(\frac{\Delta E}{k_B T}\right) \approx 1 + \frac{\Delta E}{k_B T},$$

with $k_B = 1.3806 \times \frac{10^{-23}\,\mathrm{J}}{\mathrm{K}}$ and room temperature $T \approx 300\,\mathrm{K}$. For a typical field $B \approx 0.1\,\mathrm{T}$,

$$\frac{\Delta E}{k_B T} = \frac{2\mu_B B}{k_B T} \approx \frac{2(9.274 \times 10^{-24})(0.1)}{(1.3806 \times 10^{-23})(300)} \approx 4.48 \times 10^{-4}.$$

Thus, the excess fraction of parallel spins is only about 0.0448%. In terms of percentages, if we consider 1,000,000 spins:

- Parallel: $\frac{1}{2}\left(1 + 4.48 \times 10^{-4}\right) \times 10^6 \approx 500{,}224$
- Antiparallel: $\frac{1}{2}\left(1 - 4.48 \times 10^{-4}\right) \times 10^6 \approx 499{,}776$

That is roughly 50.0224% parallel and 49.9776% antiparallel. The net magnetization is proportional to the tiny excess of about 448 spins out of a million, leading to the very small paramagnetic susceptibility $\chi \approx 1 \times 10^{-4}$ and relative permeability $\mu_r = 1 + \chi \approx 1.0001$.

*3.5.2 What Does $\mu_r = 1.4$ Mean in Terms of Net Excess?*

Our experiment yielded a measured relative permeability of $\mu_r = 1.4$ after the phase transition. This implies a magnetic susceptibility $\chi = \mu_r - 1 = 0.4$.

Magnetization is related to the applied field by $M = \chi H$. From our data in **Table.1**, for an external field $B_{\mathrm{ext}} \approx 4.2\mathrm{mT}$ (row 1), the measured contribution of the DPPH sample was $\Delta B = 1700\mu\mathrm{T} = 1.7\mathrm{mT}$. This $\Delta B$ is related to $M$ by

$$\Delta B = \mu_0 M \;\Rightarrow\; M = \frac{\Delta B}{\mu_0} = \frac{1.7 \times 10^{-3}\ \mathrm{T}}{4\pi \times \frac{10^{-7}\mathrm{H}}{\mathrm{m}}} \approx 1{,}353\ \mathrm{A/m}$$

Now we compare this to the theoretical saturation magnetization $M_{\text{sat}}$ if all unbound electrons were aligned parallel. For our 36 mg DPPH sample:

- Molar mass of DPPH $\approx 394.32$ g/mol
- Number of moles $n = \frac{0.036\ \text{g}}{394.32\ \text{g/mol}} \approx 9.14 \times 10^{-5}$ mol
- Number of unbound electrons: $N_{\text{total}} = nN_A \approx (9.14 \times 10^{-5})(6.022 \times 10^{23}) \approx 5.50 \times 10^{19}$
- Total magnetic moment for 100% alignment: $m_{\text{total}} = N_{\text{total}}\, \mu_B \approx (5.50 \times 10^{19})(9.274 \times 10^{-24}) \approx 5.10 \times 10^{-4}$ A · m$^2$
- Volume of sample (density $\approx 1$ g/cm$^3$ ): $= 0.036$ cm$^3$ $= 3.6 \times 10^{-8}$ m$^3$
- Saturation magnetization: $M_{\text{sat}} = \frac{m_{\text{total}}}{V} = \frac{5.10\times10^{-4}}{3.6\times10^{-8}} \approx 14{,}167$ A/m.

Thus, the measured magnetization $M \approx 1{,}353$ A/m is

$$\frac{M}{M_{\text{sat}}} = \frac{1{,}353}{14{,}167} \approx 0.096 = 9.6\%$$

This $\mathbf{9.6}\%$ is the net excess alignment relative to the maximum possible excess (i.e. 100% alignment would give an excess equal to the total number of spins). In other words, out of the total $N_{\text{total}}$ spins, the number that are excessively parallel (parallel minus antiparallel) is $0.096 \times N_{\text{total}}$.

*3.5.3 Converting Net Excess to Parallel/Antiparallel Percentages*

Let $N_+$be the number of parallel spins and $N_-$the number of antiparallel spins, with $N_+ + N_- = N_{\text{total}}$. The net excess is $N_+ - N_- = \Delta N$. We have

$$\frac{\Delta N}{N_{\text{total}}} = 0.096$$

Solving:

$$N_+ = \frac{N_{\text{total}} + \Delta N}{2} = \frac{N_{\text{total}}\,(1 + 0.096)}{2} = 0.548 N_{\text{total}},$$

$$N_- = \frac{N_{\text{total}} - \Delta N}{2} = \frac{N_{\text{total}}\,(1 - 0.096)}{2} = 0.452 N_{\text{total}}.$$

Therefore, the statistical distribution is approximately *54.8% parallel and 45.2% antiparallel alignment* to the external field.

This is a 9.6 percentage point difference ( $54.8 - 45.2 = 9.6$ ). In contrast, the normal paramagnetic distribution has a difference of only about 0.045 percentage points. Our experiment has increased the net alignment by a factor of roughly $9.6/0.045 \approx 213$ times, which is consistent with strongly disturbing the initial random Boltzmann distribution in our paramagnetic sample in favor of the less energy demanding ferromagnetic state of parallel aligned with the external field unbound electrons.

*3.5.4 Summary*

- Measured relative permeability: $\mu_r = 1.4$
- Implied statistical distribution of unbound electrons: approximately **55**% parallel, **45**% antiparallel.

- This represents a massive departure from the normal paramagnetic state (50.022%/49.978%) and is consistent with a genuine magnetic phase transition to a weakly ferromagnetic state.

## 4. Conclusions

We have clearly demonstrated with this novel experiment and supported by our experimental findings and data analysis that an artificially partial control of the intrinsic, random quantum spin of free electron is achievable under the special experiment conditions and preparation we described herein.

As far as we know, this is the first time ever such a research project was undertaken and successfully concluded in controlling the quantum spin of free electron to align parallel to the vectors of an external magnetic field. The novelty of our approach is that we modified the external applied magnetic field introducing a specific flux helicity to the field and characteristic helix pitch angle the field's flux, which angle is approximately equal to the known in the mathematical literature, as the Magic Angle $\theta_m \approx 54.7356°$.

Strangely enough this same "magic angle" coincides perfectly with the predicted by quantum mechanics, precession uncertainty cones radial angle. Characteristic of free electrons (see section "1.Introduction" and Fig.1) precessing under an external homogeneous B field suggesting and supported by our research here, that this "magic angle" must be an intrinsic property of the charge of the electron.

Therefore, it was inferred by the authors in this research that a modified external helical magnetic field with this exact same as possible magic angle embedded in the field could "match" this same angle intrinsic property of the charge of the electron and potentially in some extend we could gain control over the direction in space of its quantum otherwise random spin thus its intrinsic spin magnetic moment vector orientation in space.

We have put hereby our novel hypothesis under test and everything in our experiment we build around this hypothesis with the sole purpose to either disprove or prove the hypothesis correct. Fortunately, it seems from the results presented, that the later is the case with this novel experiment since we clearly demonstrated that partial control of the random quantum spin of free electrons is possible.

Furthermore, in our experiment we utilized only a limited strength up to a maximum observed of ≈38mT measured inside the DPPH material filled cavity. It remains to be seen in future experiments also undertaken by other independent research groups if the effect scales up with external field intensity gaining even more artificial control over the spin of electron inside a sample population and also if this observed novel effect remains linear or not with the increase in field strength.

Additional suggestions for future independent replication and possible more complete verification of our novel experiment, findings and subsequent conclusions include such as, angle-sweep experiment measurement under various indermediate coil winding angles from different solenoid prototypes constructed. To look up for the magic angle specificity of the novel effect observed. Other are, if posible three-axis xyz magnetometer measurements inside the 2.5 mm radius cavity and EDX (Energy Dispersive X-ray) analysis of the DPPH sample used. To prove the absence of MnZn ferrite impurities that may have contaminted the sample.

Importantly, repeating the experiment but this time utilizing a modified SG-experiment [16][17] (i.e. Stern-Gerlach experiment) setup with our proposed specific magic angle helical field in place inside a high-vacuum environment could potentially give us more definitive and conclusive results, hopefully in the near-future.

In **Fig.9** we see how the described effect would appear in the detector screen of a classical SG-experiment when modified with the proposed magic angle helical field:

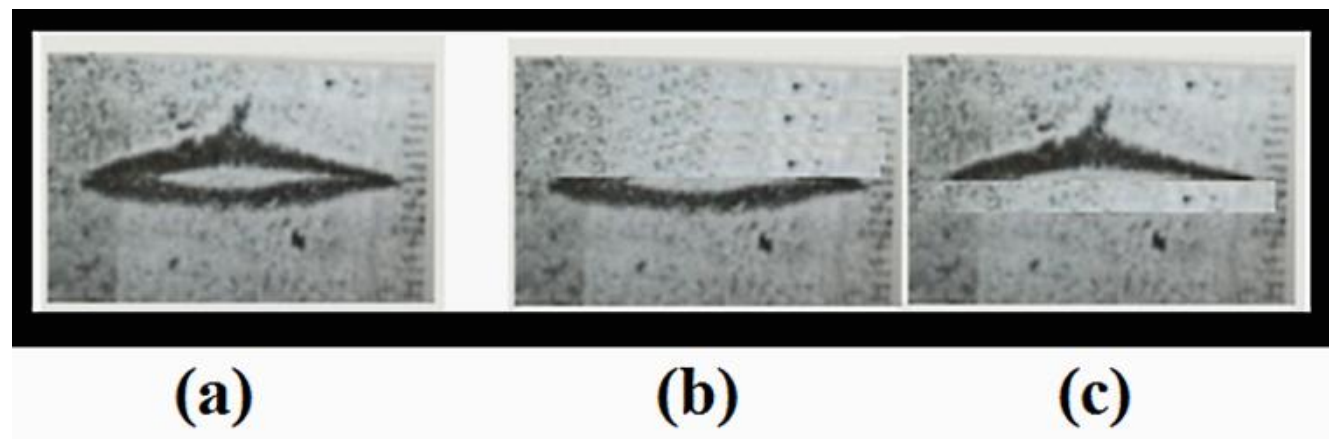

Figure 9. The observed novel effect how it would appear in a classical SG-experiment modified with the proposed novel magic angle helical field [1].

In **Fig.9(a)** we see the classical result of the SG-experiment [1] [2] with the characteristic "lips" pattern imprinted by the silver atoms hitting the detector screen and imprinting this pattern which is due to the random binary quantization spin up or spin down of the quantum spin of the atom's valance electron in space. The electrons hit only with equal 50% chance either the upper or the bottom lip but never land in the empty center area inside the lips pattern. In **Fig.9(b)&(c)** we see how the novel effect described in our experiment would accordingly appear in a modified SG-experiment using the proposed magic angle helical field instead the usual inhomogeneous field used in classical SG-magnet apparatus. Only one of the two lips the upper or the bottom lip will appear not both suggesting therefore a 100% artificial control over the quantum spin of electrons in the experiment or both the upper and bottom lip will appear imprinted but with one of the two lips being more fader than the other suggesting therefore that a partial control of the quantum spin of electrons was achieved.

Maybe one of the most fundamental implications of our research would be that it could potentially have application in quantum entanglement experiments [18] [19] [20] [21] [22] since by gaining control over the random spin of the electrons and forcing them to align by will parallel or antiparallel to the external field makes possible to assign binary bit information logic 1 and 0 to the spin up and spin down quantum states. Thus, open up the way for future quantum communication systems [23] [24].

Our research [25] if verified will significantly promote further research for the quantum control of the random spin of the electrons usually in paramagnetic materials and therefore potentially for certain materials make possible the phase transition from being paramagnetic to become ferromagnetic at room temperature using a specific design helical external magnetic field. This can have interesting applications in quantum computing and quantum communications and in general benefit quantum entanglement research and quantum electronics.

**Acknowledgments:** The authors are grateful to the peer-reviewers of the manuscript that helped to make it better.

Declarations: The authors have no relevant financial or non-financial interests to disclose.

**Data Availability Statement:** The data that supports the findings of this study are available within the article.

Supplementary material available at institutional permalink google drive repository: https://tinyurl.com/34xdt5wr.